\documentclass[12pt]{iopart} 
 
\usepackage[dvips]{graphicx} 

\newcommand{\dslash}{{\partial\!\!\!/}}     
 
\begin{document} 
 
\title[1-2-3-flavor color superconductivity in compact stars] 
{1-2-3-flavor color superconductivity in compact stars} 
 
\author{David Blaschke$^{1,2}$,  Fredrik Sandin$^{3}$ and 
Thomas Kl\"ahn$^{4}$ } 
 
\address{ 
$^{1}$ Institute for Theoretical Physics, University 
of Wroclaw, 50-204 Wroclaw, Poland\\ 
$^{2}$ Bogoliubov Lab.~
for Theoretical Physics, 
JINR Dubna, 141980 Dubna, Russia\\ 
$^{3}$ IFPA, D\'epartement AGO, Universit\'e de Li\`ege, 4000 Li\`ege, 
Belgium\\ 
$^{4}$ Physics Division, Argonne National Laboratory, Argonne, IL 60439, USA} 
\ead{blaschke@ift.uni.wroc.pl} 
 
\begin{abstract} 
We suggest a scenario where the three light quark flavors are sequentially 
deconfined under increasing pressure in cold asymmetric nuclear matter as, 
e.g., in neutron stars.  
The basis for our analysis is a chiral quark matter model of 
Nambu--Jona-Lasinio (NJL) type with diquark pairing in the single 
flavor color-spin-locking (CSL), two-flavor (2SC) and three-flavor 
color-flavor locking (CFL) channels, and a 
Dirac-Brueckner Hartree-Fock (DBHF) approach in the nuclear matter sector. 
We find that nucleon dissociation sets in at about the saturation density, 
$n_0$, when the down-quark Fermi sea is populated (d-quark dripline) due to 
the flavor asymmetry imposed by $\beta$-equilibrium and charge neutrality.  
At about $3n_0$ u-quarks appear forming a two-flavor color superconducting 
(2SC) phase, while the s-quark Fermi sea is populated only at still higher 
baryon density. 
The hybrid star sequence has a maximum mass of 2.1 M$_\odot$. 
Two- and three-flavor quark matter phases are found only in gravitationally 
unstable hybrid star solutions. 
\end{abstract} 
 
 
{\it Introduction -} 
Recent results from observations of compact star properties 
provide constraints on the nuclear equation 
of state (EoS) \cite{Klahn:2006ir}. 
In particular, the high masses $M\sim 2.0~M_\odot$ of  
compact stars in low-mass X-ray binaries, e.g., 4U 1636-536   
\cite{Barret:2005wd} and the indicated large radius $R > 12$ km of the 
isolated neutron star RX J1856.5-3754 \cite{Trumper:2003we} point  
to a stiff EoS at high densities. 
Other limits on the stiffness of the EoS comes from heavy-ion collision 
data for kaon production  
and elliptic flow  
(see \cite{Klahn:2006ir} for references). 
A key question is whether the phase transition to quark matter can occur 
inside compact stars \cite{Alford:2006vz} and if it is accompanied by 
observable signatures. 
Based on a DBHF approach in the nuclear matter sector and a chiral quark matter
model with a vector meanfield interaction a recently developed class of hybrid 
EoS  \cite{Klahn:2006iw,Blaschke:2007ri} obtained stable hybrid stars with  
masses from 1.2 M$_\odot$ up to 2.1 M$_\odot$ in accordance with  modern 
mass-radius constraints. 
Under the $\beta$-equilibrium condition in compact stars 
the chemical potentials of quarks and electrons are related by 
$\mu_d=\mu_s$ and $\mu_d=\mu_u+\mu_e$. 
The mass difference between the strange and the light quark flavors  
$m_s > m_u, m_d$ has two important consequences: (1) the down and strange 
quark densities are different so charge neutrality requires a 
finite electron density and, consequently, (2) $\mu_d>\mu_u$. 

When increasing the baryochemical potential, the d-quark chemical potential 
is the first to reach a critical value where the  
the partial density of free d-quarks becomes finite 
in a first order phase transition. 
Due to the finite value of $\mu_e$ the u-quark chemical potential  
is below the critical value and the s-quark density is zero due to 
the high s-quark mass. A {\it single-flavor} d-quark phase therefore 
forms.  
%
We discuss this phase here 
under the natural assumption that the neutralizing background is provided 
by nuclear matter, for details see \cite{Blaschke:2008br}.  
 
\begin{figure} [!ht] 
\begin{tabular}{ll} 
\includegraphics[angle=0,width=0.45\textwidth,height=0.36\textwidth]{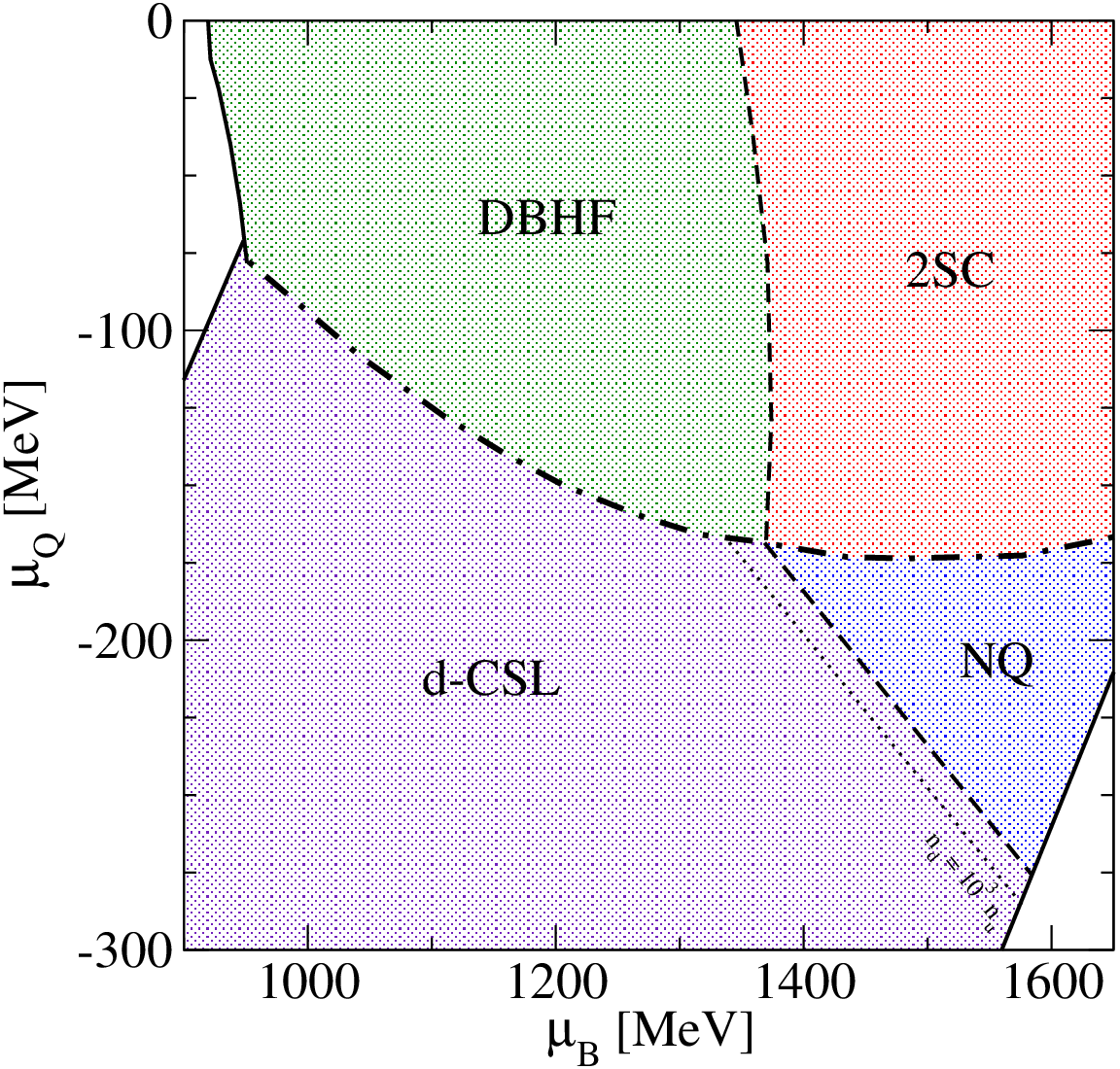}& 
\includegraphics[angle=0,width=0.45\textwidth,height=0.36\textwidth]{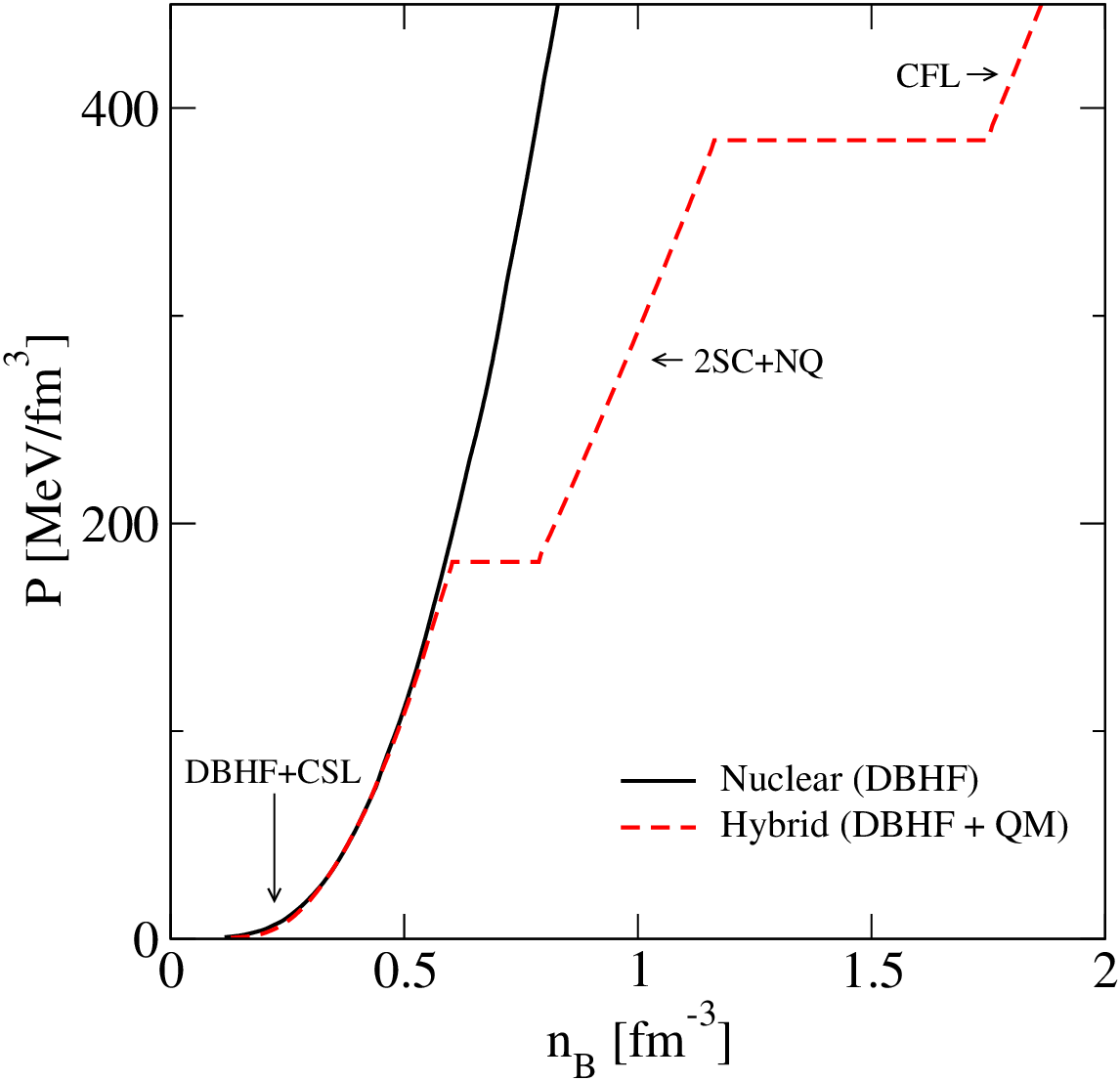} 
\end{tabular} 
\caption{Left panel: Phase diagram in the plane of baryon and charge chemical  
potential. The dash-dotted line denote the border between oppositely charged  
phases. 
Right panel: Pressure of matter in $\beta$-equilibrium as a function of the  
density.} 
    \label{f:phases} 
\end{figure} 
 
{\it Phase transition to quark matter: nucleon 
dissociation -} 
The task to describe the phase transition from nuclear matter to quark 
matter as a dissociation of three-quark bound states into their 
quark constituents (Mott transition) has not yet been solved.  
Only some aspects have been considered yet, e.g., within a nonrelativistic 
potential model \cite{Ropke:1986qs} or within the NJL model 
\cite{Lawley:2006ps}. 
We will consider a chemical equilibrium of the type 
$n+n \leftrightarrow p + 3 d$, which results in a mixed phase of nucleons 
and down quarks once the d-quark chemical potential exceeds the critical value.
This scenario is analogous to the dissociation of nuclear clusters in the 
crust of neutron stars (neutron dripline) and the effect may therefore 
be called the {\em d-quark dripline}. 
We approximate the quark and nucleon components as subphases described by 
separate models, e.g., by the DBHF approach and a three-flavor quark model  
of NJL type \cite{Blaschke:2005uj,Ruster:2005jc,Abuki:2005ms,Warringa:2005jh}. 
The partition function for the latter is given in path-integral representation 
by  
\begin{eqnarray}   
\label{Z}  
Z(T,\hat{\mu})&=&\int {\mathcal D}\bar{q}{\mathcal D}q   
\exp \left\{\int_0^\beta d\tau\int d^3x\,\left[   
        \bar{q}\left(i\dslash-\hat{m}_0+\hat{\mu}\gamma^0\right)q+  
{\mathcal L}_{\rm int}   
\right]\right\},  
\end{eqnarray}   
{\small
\begin{eqnarray}   
\label{Lint}  
{\mathcal L}_{\rm int} &=& G_S\left\{ \sum_{a=0}^{8}
        \left[(\bar{q}\tau_aq)^2 + (\bar{q}i\gamma_5\tau_a q)^2 \right]
        +\eta_{D0}\hspace{-3mm}\sum_{A=2,5,7} j_{D0,A}^\dagger  j_{D0,A} 
        +\eta_{D1}~j_{D1}^\dagger j_{D1} \right\},  \nonumber 
\end{eqnarray}   
}
where 
$\hat{\mu}=\frac{1}{3}\mu_B
+{\rm diag}_f(\frac{2}{3},-\frac{1}{3},-\frac{1}{3})\mu_Q
+\lambda_3\mu_3+\lambda_8\mu_8$ 
is the diagonal quark chemical  potential matrix  
and $\hat{m}_0={\rm diag}_f(m_u^0,m_d^0,m_s^0)$  
the current-quark mass matrix.  
For $a=0$, $\tau_0=\sqrt{2/3}~{\mathbf 1}_f$, otherwise $\tau_a$ and   
$\lambda_a$  are Gell-Mann matrices acting in flavor and color spaces,  
respectively.  
$C=i\gamma^2\gamma^0$ is the charge conjugation operator and   
$\bar{q}=q^\dagger\gamma^0$.  
The coupling strength, $G_S$, of the scalar quark-antiquark current-current 
interaction and the 3-momentum cutoff, $\Lambda$, are  
fixed by low-energy QCD phenomenology (see table I of \cite{Grigorian:2006qe}).
The relative coupling strengths of the 
spin-0 and spin-1 diquark currents, 
$j_{D0,A}=q^TiC\gamma_5\tau_A\lambda_Aq$ and  
$j_{D1}=q^TiC(\gamma_1\lambda_7+\gamma_2\lambda_5+\gamma_3\lambda_2)q$, 
are essentially free parameters,  but we restrict the discussion 
to the values from a Fierz transformation of the one-gluon exchange 
interaction, $\eta_{D0}=3/4$ and $\eta_{D1}=3/8$, see  \cite{Aguilera:2005tg}. 
The mass gaps, pairing gaps and quark matter EoS are obtained from the 
mean-field thermodynamic potential, $\Omega_{MF}=-T\ln Z_{MF}$, which 
is obtained from (\ref{Z}). 
In Fig. 1 the phase diagram and hybrid EoS are illustrated. 
The phase diagram maps the phases with lowest free energy in the 
plane of baryon chemical potential, $\mu_B$, and electric charge chemical 
potential, $\mu_Q$. 
The EoS is the path between oppositely charged phases.  
Volume fractions are determined such that the 
mixture of the two phases is neutral. 
The equations for the mass gaps and pairing gaps are solved separately 
for the CSL \cite{Aguilera:2005tg} ($\eta_{D0}=0$) and the 2SC/CFL 
\cite{Blaschke:2005uj}  ($\eta_{D1}=0$) phases, see \cite{Alford:2007xm}
for a recent review.
Each subphase is locally color neutral, and the mixture is charge neutral 
and $\beta$-equilibrated \cite{Glendenning:1992vb}. 
The discussion of surface tension and charge screening effects goes beyond the 
present exploratory investigation. It shall be performed along 
the lines of Refs. \cite{Alford:2001zr,Reddy:2004my,Voskresensky:2001jq}.
\begin{figure}[!ht] 
\begin{tabular}{ll} 
\includegraphics[angle=0,width=0.45\textwidth,height=0.4\textwidth]{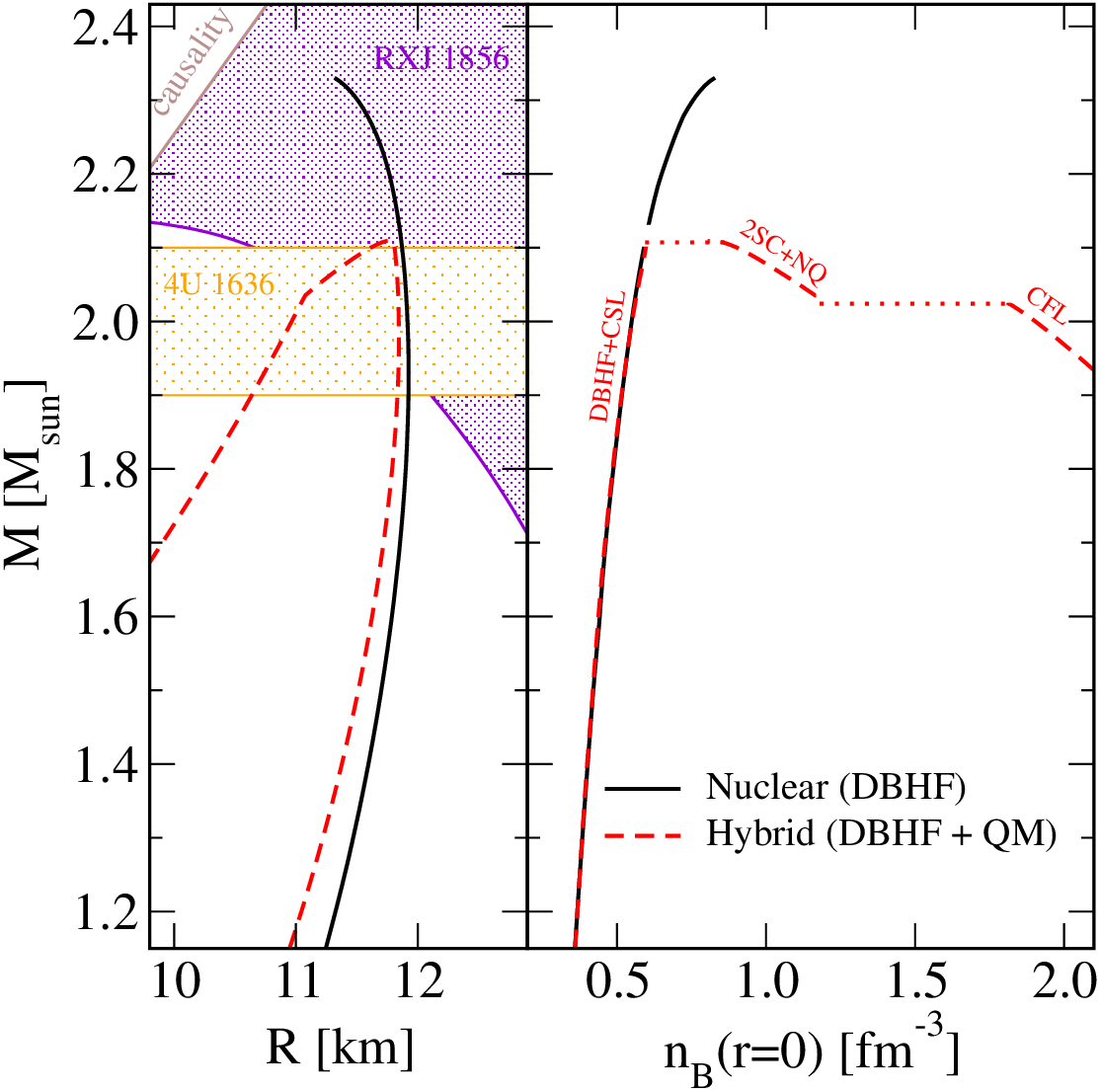}& 
\includegraphics[angle=0,,width=0.45\textwidth,height=0.4\textwidth]{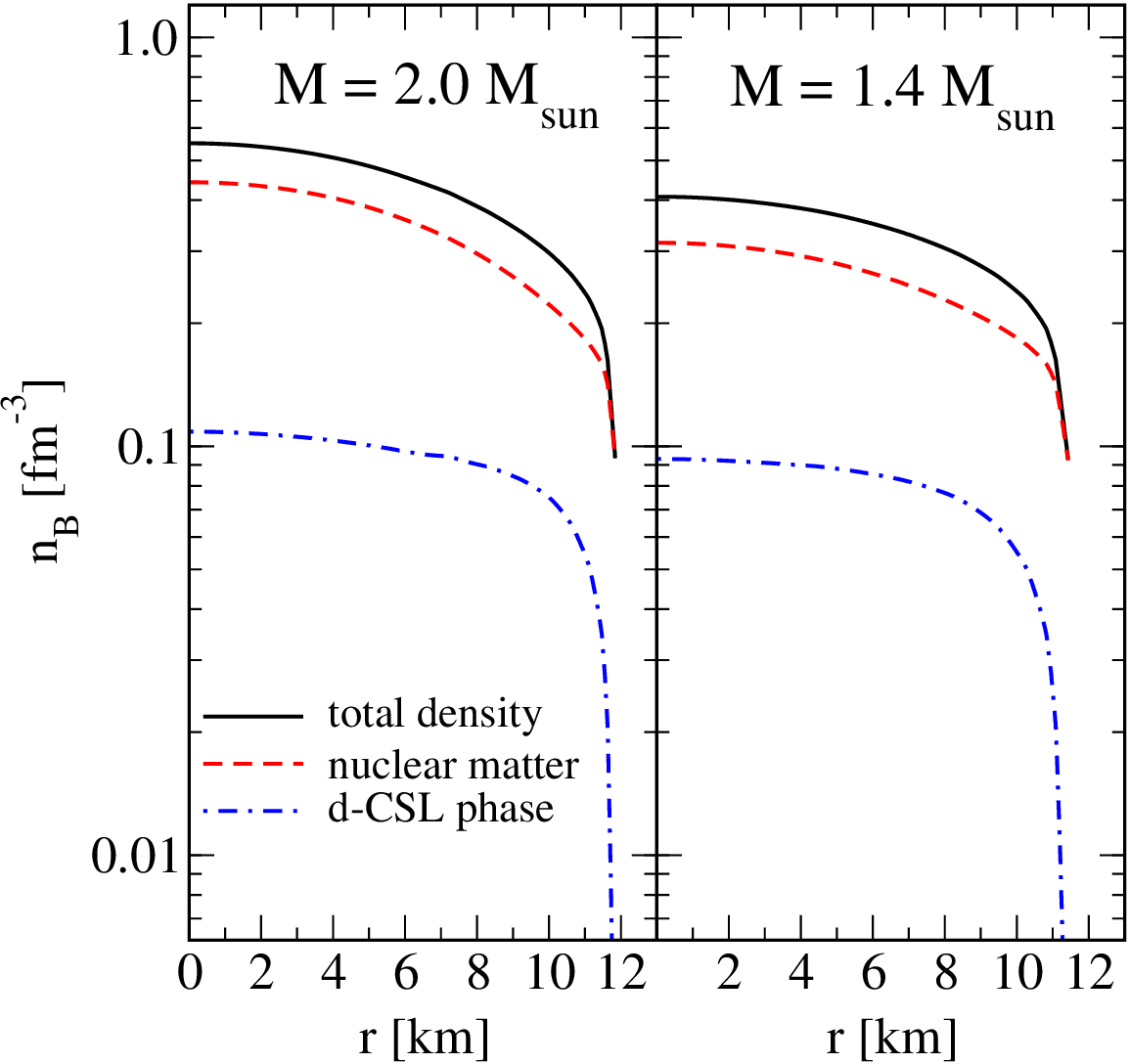} 
\end{tabular} 
\caption{Left panel: 
Compact star sequences. The phase structure of the core changes with  
increasing density, as indicated in the figure. Constraints on the mass come 
from 4U 1636 \cite{Barret:2005wd} and on the mass-radius relation from  
RX J1856 \cite{Trumper:2003we}. 
Right panel: Density profiles of two stars with masses $1.4$~M$_\odot$ and  
$2.0$~M$_\odot$. 
} 
    \label{f:config} 
\end{figure} 
 
In Fig. 2 we show compact star sequences that correspond to  
the hybrid EoS described above  
and a $\beta$-equilibrated nuclear DBHF EoS. 
We also plot density profiles of two stars with masses 
$1.4$~M$_\odot$ and $2.0$~M$_\odot$. 
It has been suggested that matter at high densities can be 
probed with analyses of the cooling behavior  
\cite{Blaschke:2006gd,Popov:2004ey,Popov:2005xa,Blaschke:1999qx} 
and the stability of rapidly rotating stars against r-modes  
\cite{Madsen:1999ci,Drago:2007iy}. 
Since the mixed d-quark CSL and nuclear matter phase extends up to the 
crust-core boundary and therefore could have significant consequences, 
we suggest further investigations to clarify the impact of this new 
phase using the probing tools mentioned above. 
  
{\it Conclusions -} 
In this contribution we have suggested a new quark-nuclear hybrid equation 
of state for compact star applications that fulfills modern mass-radius  
constraints. 
Due to isospin asymmetry, down-quarks may ``drip out'' from nucleons and  
form a single-flavor color superconducting (CSL) phase coexisting  
with nuclear matter already at the crust-core boundary in compact stars. 
The CSL phase has interesting cooling and transport properties that are in 
accordance with constraints from the thermal and rotational evolution of  
compact stars \cite{Blaschke:2007bv}.  
It remains to be investigated whether this new compact star 
composition could lead to unambiguous observational consequences. 
Eventually, it could contribute to an explanation of the puzzling superburst 
phenomenon \cite{Page:2005ky}. 
 
{\it Acknowledgements -}  
We thank  
J. Berdermann for providing results for the CSL phase and  
C. Fuchs for the DBHF EoS. 
D.B. is supported in part by the Polish Ministry of National Education  
(MENiSW). F.S. acknowledges support from the Belgian fund for scientific 
research (FNRS). T.K. is grateful for support from the Department 
of Energy, Office of Nuclear Physics, contract no.\ DE-AC02-06CH11357. 
D.B. and F.S. thank the European Science Foundation (ESF) for support from 
the CompStar programme.

\end{document}